\documentclass[11pt]{article}
\usepackage{amsmath,amssymb,color,epsfig,cite}
\usepackage{graphicx}
\usepackage{subfigure}
\usepackage{setspace}

\usepackage{amsfonts}
\def\td{\tilde}
\newcommand{\hoch}[1]{$\, ^{#1}$}

\makeatletter
\@addtoreset{equation}{section}
\makeatother

\newcommand{\be}{\begin{equation}}
\newcommand{\ee}{\end{equation}}
\newcommand{\bea}{\setlength\arraycolsep{2pt} \begin{eqnarray}}
\newcommand{\eea}{\end{eqnarray}}
\newcommand{\nn}{\nonumber}

\def\ft#1#2{{\textstyle{\frac{\scriptstyle #1}{\scriptstyle #2} } }}
\def\fft#1#2{{\frac{#1}{#2}}}
\def\CP{{{\mathbb C}{\mathbb P}}}
\def\0{{\sst{(0)}}}
\def\1{{\sst{(1)}}}
\def\2{{\sst{(2)}}}
\def\3{{\sst{(3)}}}
\def\4{{\sst{(4)}}}
\def\5{{\sst{(5)}}}
\def\6{{\sst{(6)}}}
\def\7{{\sst{(7)}}}
\def\8{{\sst{(8)}}}
\def\sst#1{{\scriptscriptstyle #1}}
\def\oneone{\rlap 1\mkern4mu{\rm l}}
\def\ep{{\epsilon}}
\def\del{{\partial}}

\def\im{{{\rm i\,}}}

\def\wtd{\widetilde}
\def\tG{{\widetilde\Gamma}}
\def\vp{{\varphi}}
\def\si{{\psi}}
\def\dlt{{\Delta}}

\begin{document}

\begin{flushright}
\hfill{UPR-1289-T\ \ \ MI-TH-1875}

\end{flushright}

\begin{center}
{\large {\bf BPS Kerr-AdS Time Machines}}

\vspace{15pt}
{\large
M. Cveti\v c\hoch{1,2\dagger}, Wei-Jian Geng\hoch{1\ddagger}, H. L\"u\hoch{3*}
and C.N. Pope\hoch{1,4,5\sharp} }

\vspace{15pt}

\hoch{1}{\it Department of Physics, Beijing Normal University, Beijing 100875, China}
\vspace{10pt}

{\hoch{2}\it Department of Physics and Astronomy,\\
University of Pennsylvania, Philadelphia, PA 19104, USA}
\vspace{10pt}

\hoch{3}{\it Department of Physics, Tianjin University, Tianjin 300350, China}

\vspace{10pt}

\hoch{4}{\it George P. \& Cynthia Woods Mitchell  Institute
for Fundamental Physics and Astronomy,\\
Texas A\&M University, College Station, TX 77843, USA}

\vspace{10pt}

\hoch{5}{\it DAMTP, Centre for Mathematical Sciences,
 Cambridge University,\\  Wilberforce Road, Cambridge CB3 OWA, UK}

\vspace{30pt}

\underline{ABSTRACT}
\end{center}

It was recently observed that Kerr-AdS metrics with negative mass
describe smooth spacetimes that have a region with
naked closed time-like curves, bounded
by a velocity of light surface. Such spacetimes are sometimes
known as time machines.
In this paper we study the BPS limit of these metrics, and find that the
mass and angular momenta become discretised.  The completeness of the
spacetime also requires that the time coordinate be periodic, with
precisely the same period as that which arises for the global AdS in which
the time machine spacetime is immersed.  For the case of equal angular
momenta in odd dimensions, we construct the Killing spinors explicitly,
and show they are consistent with the global structure.  Thus in examples
where the solution can be embedded in a gauged supergravity theory, they
will be supersymmetric.  We also compare the global structure of the
BPS AdS$_3$ time machine with the BTZ black hole, and show that the
global structure allows to have two different supersymmetric limits.

\vfill
{\footnotesize \noindent\hoch{\dagger}cvetic@physics.upenn.edu\ \ \hoch{\ddagger}gengwj@mail.bnu.edu.cn\  \ \hoch{*}mrhonglu@gmail.com\ \ \hoch{\sharp}pope@physics.tamu.edu}

\thispagestyle{empty}

\pagebreak

\tableofcontents
\addtocontents{toc}{\protect\setcounter{tocdepth}{2}}


\newpage

\section{Introduction}

The Kerr metric \cite{Kerr:1963ud} is arguably the most important exact
vacuum solution in Einstein's theory of General Relativity.  Over the
years, the solution has been generalised to include the cosmological
constant and also to higher dimensions \cite{Carter:1968ks,Carter:1973rla,
Myers:1986un,Hawking:1998kw,Gibbons:2004uw,Gibbons:2004js}. These metrics
are asymptotic to Minkowski, de Sitter (dS) or anti-de Sitter (AdS)
spacetimes, depending on the cosmological constant. They carry mass ($M$)
and angular momenta ($J_i$) as conserved quantities.  For a given set of
angular momenta, provided that the mass is sufficiently large, the metrics
describe rotating black holes.  Such a rotating black hole contains
closed-time-like curves (CTCs), surrounded by a velocity of light
surface (VLS), which is typically referred to as a time machine.\footnote{This
 should be distinguished from what happens in the G\"odel-like or
G\"odel-type universe \cite{Godel:1949ga}, where the normal region of the
spacetime is surrounded by the VLS, outside of which lie the naked CTCs.
In this paper we shall not be concerned with naked CTCs of the G\"odel style.
We also clarify that although the time coordinate in global AdS$_D$ embedded
as a hyperboloid in $\mathbb{R}^{2,D-1}$ is periodic, AdS$_D$ is not referred
to as a time machine since it has no VLS.}  In a rotating black hole, the
time machine is hidden inside the black hole event horizon.

  If the
black hole is over-rotating, the time machine can extend outside the
horizon.  For example, it was demonstrated, for a supersymmetric
charged black hole with equal angular momenta in five dimensions
\cite{Breckenridge:1996is}, that in the over-rotating situation the boundary
of the time machine lies outside the horizon and so it becomes naked \cite{Gibbons:1999uv}.
(See also \cite{Gauntlett:1998fz,Herdeiro:2000ap,Herdeiro:2002ft}.) An
examination of geodesics showed that they could not penetrate the horizon,
and hence the spacetime configuration is called a repulson
\cite{Gibbons:1999uv}. (See also \cite{Diemer:2013fza}.)
In fact the ``horizon'' become a Euclidean Killing
horizon that can induce a conical singularity unless the real time
coordinate becomes periodic with some specific period,
in which case the spacetime configuration is smooth and geodesically complete \cite{Cvetic:2005zi}.

    Recently, it was observed \cite{Feng:2016dbw} that for general
odd dimensions, Kerr and Kerr-AdS
metrics can also generally extend onto smooth time machines when mass
is negative, provided that all the angular momenta are non-vanishing.
For Kerr-AdS metrics, there exist special points in the parameter space of
the mass and charges, namely when the BPS condition
\be
M=\sum_i g J_i\,,\label{BPScons}
\ee
holds, where $1/g$ is the ``radius'' of the AdS spacetime in which the
solution is immersed.  When this BPS condition is satisfied, the spacetime
admits a Killing spinor.
The BPS condition was studied in \cite{Cvetic:2005nc} for the five-dimensional
Kerr-AdS black hole, and the Killing spinors were constructed in the
case where the two equal angular momenta were equal.  The BPS limit of the
Kerr-AdS metric no longer describes a black hole, however, since the
singularity is no longer shielded by an horizon. Interestingly, if one
Euclideanises the spacetime and takes the cosmological constant to be
positive the BPS Kerr-dS becomes an Einstein-Sasaki metric,  which can
smoothly extend onto a complete, compact manifold for appropriate
discretised values of the metric parameters \cite{Cvetic:2005ft,Cvetic:2005vk}.
This generalises an earlier construction of smooth Einstein-Sasaki spaces
in \cite{Gauntlett:2004yd}.

In this paper, we shall remain in Lorentzian signature and consider the case
with a negative cosmological constant, but now we consider the BPS
Kerr-AdS metrics where the mass is taken to be negative. Now, unlike
the example considered in \cite{Cvetic:2005nc} where the mass was
assumed to be positive, this yields a smooth time-machine spacetime.
BPS time machines have been constructed previously in literature,
typically having positive mass and with additional electric
charges\cite{Gibbons:1999uv,Klemm:2000vn,Klemm:2000gh,Cvetic:2005zi}. In this
paper, we focus on the pure gravity BPS Kerr-AdS metrics. We analyse the
global structure and find that the metrics can become smooth,
provided that the mass is negative. The completeness of the
spacetime requires that the asymptotic Lorenzian time coordinate be
periodic, with period precisely equal to that of the time
coordinate in the global AdS in which the spacetime is immersed.
Furthermore, the mass and angular momenta become discretised,
in a manner analogous to the discretisation of the parameters in
the Einstein-Sasaki spaces, even though the spacetimes we are considering
here are non-compact. For Kerr-AdS metrics with equal angular momenta in
odd dimensions, we construct the Killing spinors in the BPS limit
explicitly, and show that they are compatible with the global structure
required for the completeness of the spacetime.  Thus in dimensions
where the solution can be embedded within a supergravity theory, it
will be superymmetric.

The paper is organised as follows.  In section 2, we begin by
reviewing the time machine spacetimes that were obtained in
\cite{Feng:2016dbw} from $D=(2n+1)$-dimensional
Kerr-AdS spacetimes with equal angular
momenta, by taking the mass to be negative, and we describe their
BPS limits.  We give an explicit construction of the Killing spinors in
the BPS spacetimes, showing how they can be obtained by making use of
the gauge-covariantly constant spinors that exist in the underlying
$\CP^{n-1}$ spaces that form the bases of the $(2n-1)$-dimensional
spherical surfaces in the spacetimes.  We also study the restrictions
on the metric parameters that result from requiring completeness of
the spacetimes, resulting from the compatibility conditions for periodicities
at the various degenerate surfaces. These restrictions imply that
mass and angular momentum must be rational multiples of a basic unit.
They also imply that the time coordinate must be periodic, with exactly
the periodicity of the time coordinate in the global AdS spacetime in
which the time machine is immersed.

  In section 3 we consider the case of even-dimensional spacetimes,
showing that Kerr-AdS metrics with equal angular momenta can give rise in
the BPS limit to metrics describing foliations of the previously discussed
odd-dimensional time machines.  In section 4 we discuss the analogous
odd and even-dimensional BPS limits of Kerr-AdS metrics with general,
unequal, angular momenta.  Again these give rise to time machines
if the mass is taken to be negative, and we analyse the restrictions on
the metric parameters to ensure global completeness of the spacetime
manifolds.  Again, the mass and the angular momenta are discretised, in
the sense that they are constrained to be certain rational multiples of
a basic unit.

   In section 5 we discuss the special case of three dimensions.  Here,
the Kerr-AdS metric is necessarily locally isomorphic to AdS$_3$, and
thus it is also locally isomorphic to the BTZ black hole \cite{Banados:1992wn}.
We study the relation between the time machine and the BTZ spacetimes, and compare their
Killing spinors in the respective BPS limits.  Interestingly, the
limits are different, but in each case the Killing spinors are compatible
with the global structures.

   Finally, after our conclusions, we include two appendices.  Appendix A
gives an explicit construction of the gauge-covariantly spinors in
the complex projective spaces, employing an iterative construction of
$\CP^n$ in terms of $\CP^{n-1}$ that was given in \cite{Hoxha:2000jf}. We use
these gauge-covariantly constant spinors in the construction of Killing
spinors in section 2.  Appendix B contains some results relating the various
vectors and tensors that can be built from Killing-spinor bilinears.

\section{Equal angular momenta in $D=2n+1$}\label{equaloddsec}

\subsection{Kerr black holes and time machines}

      We begin with the Kerr-AdS metrics in $D=2n+1$ dimensions with all $n$
angular momenta set equal.  The metric, satisfying $R_{\mu\nu}=
-(D-1) g^2 g_{\mu\nu}$, contains two integration constants
$(m, a)$, and it is given by \cite{Gibbons:2004ai}
\bea
ds_{2n+1}^2 &=& - \fft{1 + g^2 r^2}{\Xi} dt^2 + \fft{U dr^2}{V-2m} + \fft{r^2 + a^2}{\Xi}
(\sigma^2 + d\Sigma_{n-1}^2) + \fft{2m}{U\Xi^2}(dt - a \sigma)^2\,,\nn\\
\sigma &=& d\psi +A\,,\qquad U= (r^2 + a^2)^{n-1}\,,\qquad
V=\fft{1}{r^2} (1+ g^2 r^2) (r^2 + a^2)^n\,,\label{oddequalrot}
\eea
where $\Xi=1-a^2 g^2$, and $d\Sigma_{n-1}^2$ is the standard Fubini-Study
metric on $\mathbb {CP}^{n-1}$.  There is circle, parameterised by the
coordinate $\psi$ with period $2\pi$, which is fibred over the
$\mathbb {CP}^{n-1}$ base, and $\sigma$ is the 1-form on the fibres, given by
$\sigma=d\psi + A$ with $dA=2 J$ where $J$ is the K\"ahler form
on $\mathbb {CP}^{n-1}$.
The terms $(\sigma^2 + d\Sigma_{n-1}^2)$ in the metric are nothing but the
metric on the unit round sphere $S^{2n-1}$, with $R^i_j= (n-1) \delta^i_j$.
The metric (\ref{oddequalrot} is asymptotic to anti-de Sitter
spacetime with radius $\ell=1/g$.

    The mass and the (equal) angular momenta are given by
\be
M=\fft{m(2n-\Xi){\cal A}_{2n-1}}{8\pi \Xi^{n+1}}\,,
\qquad J= \fft{ma {\cal A}_{2n-1}}{4\pi \Xi^{n+1}}\,,
\ee
where ${\cal A}_{k}$ is the volume of a unit round $S^k$, given by
\be
{\cal A}_k = \fft{2\pi^{\fft12(k+1)}}{\Gamma[\fft12(k+1)]}\,.
\ee
It will be helpful to make a coordinate transformation and a
redefinition of the integration constants to replace $(m,a)$ by $(\mu,\nu)$,
as follows:
\be
\fft{r^2+a^2}{\Xi}\rightarrow r^2\,,\qquad
a=\sqrt{\fft{\nu}{\mu}}\,,\qquad m=\ft12\mu\Big(1 - \fft{\nu}{\mu}g^2\Big)^{n+1}\,.\label{amn}
\ee
The metric (\ref{oddequalrot}) becomes \cite{Feng:2016dbw}
\bea
ds_{2n+1}^2 &=& \fft{dr^2}{f} - \fft{f}{W} dt^2 +
r^2 W (\sigma + \omega dt)^2
+ r^2 d\Sigma_{n-1}^2\,,\cr
f &=& (1+g^2 r^2) W - \fft{\mu}{r^{2(n-1)}}\,,\qquad
W=1 + \fft{\nu}{r^{2n}}\,,\qquad
\omega = -\fft{\sqrt{\mu\nu}}{r^{2n}+\nu}\, dt\,.\label{oddequalrot2}
\eea
The mass and angular momenta become
\be
M=\fft{{\cal A}_{2n-1}}{16\pi} ((2n-1)\mu + g^2 \nu)\,,\qquad
J= -\fft{{\cal A}_{2n-1}}{8\pi} \sqrt{\mu\nu}\,.\label{odddmj}
\ee
The metric (\ref{oddequalrot2}) describes a rotating black hole if
$\mu$ and $\nu$ are both positive, and a time machine if
$\mu$ and $\nu$ are both negative \cite{Feng:2016dbw}, as we shall review later.

\subsection{BPS limits}

    Under certain conditions the metric (\ref{oddequalrot2}) will admit a
Killing spinor, obeying the equation
\be
\nabla_\mu\epsilon  + \fft{1}{2}\, g\, \Gamma_\mu\epsilon=0\,.
\label{killingspinor}
\ee
A necessary condition for this to occur is that the BPS condition
on the mass and angular momentum, namely
\be
M=n g J\,,
\ee
should hold.  This implies that
\be
\mu=g^2 \nu\,,\qquad \hbox{or}\qquad \mu = \fft{g^2\nu}{(2n-1)^2}\,.
\label{bps12}
\ee
The these two conditions correspond to $ag=$ (and hence $\Xi=0$) or
$a g=2n-1$ respectively. However, as we shall see,
only the first of these cases gives a solution admitting a Killing spinor.

  In AdS itself (i.e. $\mu=0$ and $\nu=0$), the Killing vectors
\be
K_\pm =\fft{\partial}{\partial t} \pm g \fft{\partial }{\partial \psi}\,,
\label{killingvector}
\ee
have the property that $g_{\mu\nu}\, K_\pm^\mu K_\pm^\nu=-1$, and in fact
they can each be expressed in the
form $K_\pm^\mu= \bar\epsilon_\pm \Gamma^\mu\epsilon_\pm$, where
each of $\epsilon_\pm$ is one of
the Killing spinors of the AdS spacetime.  We expect that if the BPS
spacetime where $\mu$ and $\nu$ are non-zero, obeying one or other of
the conditions in (\ref{bps12}), does admit a Killing spinor, then it should
be such that it limits to one of the aforementioned AdS Killing spinors in the
limit where $\mu$ and $\nu$ go to zero.  This means that if the BPS
spacetime admits a Killing spinor, the norm $K^\mu K_\mu$ should be
manifestly negative (see \cite{Cvetic:2005zi} for a discussion of this).
For the two cases in (\ref{bps12}) we find
\bea
\mu=g^2 \nu:\qquad&& g_{\mu\nu}\, K_+^\mu K_+^\nu =-1\,,\label{truebps}\\
\mu = \fft{g^2\nu}{(2n-1)^2}:\qquad && g_{\mu\nu}\, K_+^\mu K_+^\nu
   = -1 + \fft{n^2 g^2 \nu}{(2n-1)^2 r^{2n-2}}\,,
\label{fakebps}
\eea
where $K_+$ is defined in (\ref{killingvector}).
This indicates that (\ref{truebps}) gives rise to a true BPS limit, in the
sense that the $K_+$ Killing vector (but not $K_-$) admits a
spinorial square root, whereas for (\ref{fakebps}) it does not (nor does
$K_-$).

   For positive
$\mu=g^2\nu$, the metric has a curvature power-law naked singularity
at $r=0$.   We shall thus focus on the case when $\mu=g^2 \nu$ is
negative.  Defining $\nu=-\alpha$, the metric becomes
\bea
ds^2 &=& -\fft{f}{W}\, dt^2 + \fft{dr^2}{f} + r^2 W\, (d\psi+A + \omega dt)^2
  + r^2\, d\Sigma_{n-1}^2\,,\label{tmmet}\\
f&=& g^2 r^2 +W\,,\qquad W=1- \fft{\alpha}{r^{2n}}\,,\qquad
\omega= \fft{\alpha g}{W r^{2n}}\,,\nn
\eea
We have made the specific choice for the sign of $\sqrt{\mu\nu}\rightarrow
\sqrt{\nu^2g^2} =\mu g= -\alpha g$ when sending $\mu=\nu g$ negative,
and with this choice, the
Killing vector admitting the spinorial square root is again given by
(\ref{killingvector}) with the plus sign choice, for which we now define
\be
K=\fft{\del}{\del t} + g \fft{\del}{\del\psi}\,.\label{kvplus}
\ee

The mass and angular momentum are given by
\be
M=-\fft{ng^2\alpha}{8\pi} {\cal A}_{2n-1}\,,
\qquad J=\fft{g\alpha}{8\pi}{\cal A}_{2n-1}\label{tmMJ}
\ee
(recall that we have made the sign choice that $\sqrt{\mu\nu}\rightarrow
  + \alpha g$ when sending $\mu$ and $\nu$ negative).
The metric has a power-law curvature singularity at $r=0$, but there is a
Euclidean Killing horizon at $r=r_0>0$ for which $f(r_0)=0$.  Thus we have
\be
\alpha = (1 + g^2 r_0^2) r_0^{2n}\,.\label{alphar0}
\ee
The absence of a conical singularity at $r=r_0$
requires that the degenerate Killing
vector
\be
\ell = \fft{1}{n + (n+1) g^2 r_0^2}\Big( g r_0^2 \fft{\partial}{\partial t} +
(1 + g^2 r_0^2) \fft{\partial}{\partial\psi}\Big)\,,\label{kv1}
\ee
must generate a $2\pi$ period.  As we shall discuss later, this implies that
the $t$ coordinate must be periodically identified.
Note that we have scaled the Killing vector
so that the corresponding Euclidean surface gravity is precisely unity.

    Defining a radius $r_*\equiv \alpha^{\ft{1}{2n}}$,
we see that $g_{\psi\psi}<0$ in the region
\be
r_0 < r < r_*\,,\label{tmregion}
\ee
and thus $\psi$ is the time coordinate in this region.  (The VLS is located at $r=r_*$ where $g_{\psi\psi}=0$.) Since $\psi$ is
periodic, with perod $\psi$ as stated earlier, it follows that there are
closed timelike curves in the region defined by (\ref{tmregion}).  This
situation
is commonly described as a time machine (see \cite{Cvetic:2005zi} for a
more detailed discussion).

Finally, it is worth pointing out that in the case $\mu=g^2\, \nu$,
for which there is a Killing spinor, the corresponding
metric (\ref{tmmet}) can be
expressed, after we make a coordinate change $\psi\rightarrow \psi - g\, t$,
as a time bundle over a $D=2n$ dimensional space:
\be
ds_{2n+1}^2 = - \Big(dt + g r^2 (d\psi + A)\Big)^2 +
\fft{dr^2}{f} + r^2 \Big(f\, (d\psi + A)^2 + d\Sigma_{n-1}^2\Big)\,.
\ee
The length of the time fibre is constant, and the base is a
$2n$-dimensional Einstein-K\"ahler metric.  In fact this is
Lorentzian version of the situation in an Einstein-Sasaki space,
which can be written, at least locally, as a constant-length
circle fibration over an Einstein-K\"ahler base space.

\subsection{Killing spinors}\label{kssec}

   Here, we construct the Killing spinor $\eta$ in the $(2n+1)$-dimensional
BPS time machine with equal angular momenta, whose metric is given by
(\ref{tmmet}), obeying
\be
\nabla_a \eta + \ft12 g \Gamma_a\eta=0\,.\label{kseqn}
\ee
We shall make use of the fact
that $\CP^{n-1}$ admits a gauge-covariantly constant spinor $\xi$ satisfying
\be
\wtd D_i\, \xi + \fft{\im n}{2}\, A_i\, \xi=0\,,
\ee
where $\wtd D= \td d + \ft14 \td\omega^{ij}\, \tG_{ij}$ is the spinor-covariant
exterior derivative and $\wtd D= \td e^i \wtd D_i$, with $\tG_i$
being the Dirac matrices and $\td e^i$ denoting
a vielbein basis for $\CP^{n-1}$.\footnote{We use $\td d$ to denote
the standard exterior derivative in the $(2n-2)$-dimensional
$\CP^{n-1}$ space in order to distinguish it from $d$ which is the
exterior derivative in the full $(2n+1)$-dimensional space-time.}
With an appropriate choice of basis for
the Dirac matrices one can easily establish that $\xi$ obeys
\be
J^{ij} \tG_{ij}\, \xi = -2 \im (n-1) \xi\,,\qquad \tG_*\, \xi=\xi\,,
\ee
where $\tG_*$ denotes the chirality operator on $\CP^{n-1}$.  (We give an
iterative construction of the gauge-covariantly constant spinor $\xi$ in
appendix A.)

   We introduce the vielbein basis $e^a$ for (\ref{tmmet}), with
\be
e^0= u dt\,,\qquad e^1= \fft{dr}{v}\,,\qquad e^2=h(d\psi +A + \omega dt)\,,
\qquad e^i= r\td e^i\,,\label{vielbein}
\ee
where
\be
u= \sqrt{\fft{f}{W}}\,,\qquad v=\sqrt{f}\,,\qquad
   h= r \sqrt W\,.
\ee
The inverse vielbein $E_a$ is given by
\be
E_0=\fft1{u}\, \Big(\fft{\del}{\del t} -
                         \omega\, \fft{\del}{\del\psi}\Big)\,,\qquad
E_1= v\fft{\del}{\del r}\,,\qquad
  E_2 = \fft1{h}\, \fft{\del}{\del\psi}\,,\qquad
  E_i= \fft1{r}\, \Big( \wtd E_i - A_i\, \fft{\del}{\del\psi}\Big)\,,
\ee
where $\wtd E_i$ is the inverse vielbein for $\CP^{n-1}$.
The torsion-free spin connection $\omega^{ab}$
for the vielbein (\ref{vielbein}) is easily calculated,
leading to the spinor-covariant exterior derivative $D=d+\ft14 \omega^{ab}\,
\Gamma_{ab}$ given by
\bea
D &=& d+ e^0\, \Big(\fft{u' v}{2u}\, \Gamma_{01}
              -\fft{h \omega' v}{4 u}\, \Gamma_{12} \Big)
- e^1\, \fft{h\omega' v}{4u}\, \Gamma_{02} -
   e^2\, \Big(\fft{h' v}{2h}\, \Gamma_{12} +\fft{h \omega' v}{4u} \Gamma_{01}
  + \fft{h}{4 r^2}\,
   J^{ij}\, \Gamma_{ij}\Big) \nn\\
&&  -e^i\, \Big( \fft{v}{2r}\, \Gamma_{1i} + \fft{h}{2 r^2}\,
   J_i{}^j\, \Gamma_{2j}\Big) + \ft14 \td\omega^{ij}\Gamma_{ij}\,.
\label{tmspincov}
\eea

   Writing the $2n+1)$-dimensional Lorentz indices as
$a=(\alpha,i)$ with $\alpha=0,1,2$,
we may decompose the $(2n+1)$-dimensional Dirac matrices
in the form
\be
\Gamma_\alpha = \gamma_\alpha\otimes \tG_*\,,\qquad
   \Gamma_i = \oneone \otimes \tG_i\,,
\ee
where $\gamma_\alpha$ are $2\times 2$ Dirac matrices, which we take to be
\be
\gamma_0= \begin{pmatrix} 0& \ 1 \\ -1 & \ 0\end{pmatrix}\,,\qquad
\gamma_1= \begin{pmatrix} 0\ & \ 1  \\ 1 \ & \  0\end{pmatrix}\,,\qquad
\gamma_2= \begin{pmatrix} 1&\ 0 \\ 0 & \ -1\end{pmatrix}\,.\label{2Dirac}
\ee
It then follows that the spinor-covariant exterior derivative
(\ref{tmspincov}) is given by
\bea
D&=& \hat d\otimes \oneone + \oneone\otimes \wtd D +
    e^0\, \Big(\fft{u' v}{2u}\, \gamma_{01} -
              \fft{h \omega' v}{4 u}\, \gamma_{12} \Big)\otimes\oneone
- e^1\, \fft{h\omega' v}{4u}\,\gamma_{02}\otimes\oneone\nn\\
&& -e^2\, \Big(\fft{h' v}{2h}\, \gamma_{12}\otimes \oneone
  +  \fft{h\omega' v}{4u}\, \gamma_{01}\otimes \oneone +\fft{h}{4 r^2}\,
   J^{ij}\, \oneone\otimes\tG_{ij}\Big)\nn\\
&&
  -e^i\, \Big( \fft{v}{2r}\, \gamma_{1}\otimes\tG_*\tG_i + \fft{h}{2 r^2}\,
   J_i{}^j\, \gamma_{2}\otimes \tG_*\tG_j\Big)\,,
\eea
where $\wtd D$ is the spinor-covariant exterior derivative on $\CP^{n-1}$
that we introduced earlier, and $\hat d$ denotes the standard exterior
derivative in the three directions orthogonal to $\CP^{n-1}$, i.e.
$d=\hat d+ \td d =e^a\, E_a$ with
\bea
\hat d &=& e^\alpha\, E_\alpha = e^0\, \Big(\fft1{u}\, \fft{\del}{\del t}
   - \fft{\omega}{u}\, \fft{\del}{\del\psi}\Big) +
   e^1\, v \fft{\del}{\del r} +e^2\,  \fft1{h}\, \fft{\del}{\del\psi}\,,\\
\tilde d &=& e^i\, E_i =
  e^i\, \fft1{r}\,\Big( \wtd E_i - A_i\,\fft{\del}{\del\psi}
\Big)\,.
\eea

   With these preliminaries, it is now straightforward to obtain the
equations for the Killing spinor $\eta$ in the $(2n+1)$-dimensional
spacetime, satisfying (\ref{kseqn}). It takes the form
\be
  \eta = \epsilon\otimes \xi\,,
\ee
where $\xi$ is the gauge-covariantly constant spinor on $\CP^{n-1}$ that
we introduced earlier.  After further straightforward computations, we
find that the 2-component spinor $\epsilon$ is given by
\be
 \epsilon = \fft1{\sqrt2}\,
W^{-\ft14}\, \exp\big(-\ft12 \im g t -\ft12 \im n  \psi\big)\,
   \begin{pmatrix} (g r + \im\, \sqrt{W})^{\ft12}\\
                   -(g r - \im\, \sqrt{W})^{\ft12} \end{pmatrix}\,.\label{ks}
\ee

   We may now straightforwardly verify that the the Killing vector
(\ref{kvplus}) may be written in terms of the Killing
spinor $\eta$ as
\be
K^a = \bar\eta \Gamma^a\eta\,.\label{spinorialsqrt}
\ee

\subsection{Global considerations and discretisation of parameters}

  The discussion in this section is closely analogous to that in
\cite{Cvetic:2005ft,Cvetic:2005vk}, where the global structure of Einstein-Sasaki spaces
was studied.  We begin by defining the Killing vectors
\be
\ell_0 = \fft{1}{g} \fft{\partial}{\partial t}\,,
  \qquad \ell_1 = \fft{\partial}{\partial \psi}\,,
\ee
where we have included a $1/g$ in the definition of $\ell_0$ in order to make
it dimensionless.  $\ell_1$ generates a $2\pi$ period.
It follows from (\ref{kv1}) that
\be
g^2 r_0^2\, \ell_0 = \big[ n + (n+1) g^2 r_0^2\big] \, \ell -
   (1 + g^2 r_0^2)\, \ell_1\,.\label{l0l1l2rel}
\ee
Since $\ell$ and $\ell_1$ both generate periodic translations by $2\pi$, the
ratio of their coefficients must be rational, since otherwise one there would
be identifications in the time direction, generated by $\ell_0$, of
arbitrarily close points in the spacetime manifold.
Hence $g^2 r_0^2$ must be rational, which we shall write as
$g^2 r_0^2 =p/\tilde q$,
for coprime integers $p$ and $\tilde q$.  Consequently (\ref{l0l1l2rel}) can be
written as
\be
p \ell_0 = q\, \ell + q_1\, \ell_1\,,\label{l0l1l2rel2}
\ee
where the integers $q$ and $q_1$ are given by
\be
q= (n+1) p + n \tilde q\,,\qquad q_1 = -(p+\tilde q)\,.\label{q1q2def}
\ee
Note that the set of integers $\{p,q,q_1\}$ are necessarily coprime, since
$p$ and $\tilde q$ are coprime.

   It is straightforward also to see from (\ref{q1q2def}) that since
$p$ and $\tilde q$ are coprime, it must also be the case that $q$ and $q_1$
are coprime.  It then follows from (\ref{l0l1l2rel2}) that $\ell_0$
generates a smallest translation period of $2\pi$, and hence that
$g t$ has period $2\pi$.  Interestingly, this is precisely the same as the
period of the time coordinate in a global AdS with radius $g^{-1}$.  Thus
the periodicity of $t$ that is required in order to eliminate the
conical singularity at the Euclidean Killing horizon at $r=r_0$ is exactly
the same as the time periodicity of the embedding AdS spacetime itself.
Consequently, the Killing spinor (\ref{ks}) is consistent with the
global structure of the time machine spacetime, and hence the solution would be
supersymmetric if it can be embedded in a gauged supergravity.

  The fact that $g^2 r_0^2 = p/\tilde q$ is rational implies that the possible
masses (and angular momenta) for the BPS time-machine spacetimes are
discretised.  From (\ref{tmMJ}) and (\ref{alphar0}), we have
\be
M=-n J=
- \fft{n {\cal A}_{n-1}}{8\pi\, g^{2n-2}}\, \Big(1+\fft{p}{\tilde q}\Big)
   \Big(\fft{p}{\tilde q}\Big)^n\,.
\ee

\section{Equal angular momenta in $D=2n$}

The Kerr-AdS metrics in even $D=2n$ dimensions with all equal angular momenta can be expressed as \cite{Gibbons:2004ai}
\bea
ds^2 &=& - \fft{\Delta_\theta (1 + g^2 r^2)}{\Xi} dt^2 +
\fft{U dr^2}{V-2m} + \fft{\rho^2 d\theta^2}{\Delta_\theta} +
\fft{r^2 + a^2}{\Xi}\sin^2\theta[(d\psi + A)^2 + d\Sigma_{n-2}^2]\cr
&& + \fft{2m}{U\Xi^2}[\Delta_\theta\, dt - a \sin^2\theta (d\psi +A)]^2\,,
\eea
where
\bea
U &=& \fft{\rho^2 (r^2 + a^2)^{n-2}}{r}\,,\qquad
V=\fft{1}{r} (1 + g^2 r^2) (r^2 + a^2)^{n-1}\cr
\Delta_\theta &=& 1 - a^2 g^2 \cos^2\theta\,,\qquad
\rho^2 = r^2 + a^2 \cos^2\theta\,,\qquad
\Xi=1-a^2 g^2\,.
\eea
The mass and the (equal) angular momenta are \cite{Gibbons:2004ai}
\be
M=\fft{n\,m\,{\cal A}_{D-2}}{4\pi \Xi^{n}}\,,\qquad
J=\fft{m a {\cal A}_{D-2}}{4\pi \Xi^{n}}\,.
\ee
The BPS limit $M=n g J$ implies that $a g=1$ and hence $\Xi\rightarrow 0$.  This requires that
\be
m\sim \Xi^n\rightarrow 0\,,
\ee
so that $M$ and $J$ remain finite.  In this limit, for the metric to be real and the coordinate $\theta$ to be spacelike, we need make the coordinate transformation
\be
\theta \rightarrow \ft12\pi - {\rm i}\theta\qquad r^2 + a^2 \rightarrow \Xi r^2\rightarrow 0\,.
\ee
After some algebra we end up
\be
ds_{2n}^2 = g^{-2} d\theta^2 + \cosh^2\theta\, ds_{2n-1}^2\,.
\ee
where $ds_{2n-1}^2$ is the time machine metric obtained earlier for odd dimensions with all equal angular momenta.  In deriving this, we need to further redefine the scaled $m$ as
\be
m\rightarrow {\rm i}\fft{m}{g}\,.
\ee
The origin of this is that in the $(V-2m)$ factor, there is a term of $2m r$.

\section{General non-equal angular momenta}

In this section, we consider the BPS limit of general Kerr-AdS black holes with general angular momenta.

\subsection{$D=5$}

The Kerr-AdS metric in five dimensions was constructed in \cite{Hawking:1998kw}, given by
\bea
ds^2_5 =&& -\frac{\Delta_r}{\rho^2}[dt - \frac{a \sin^2 \theta}{\Xi_a} d\phi_1 - \frac{b \cos^2\theta}{\Xi_b} d\phi_2]^2 + \frac{\Delta_\theta \sin^2 \theta}{\rho^2}[a dt - \frac{r^2 +a^2}{\Xi_a} d\phi_1]^2 \cr
        &&+\frac{\Delta_\theta \cos^2\theta}{\rho^2}[b\, dt - \frac{r^2 +b^2}{\Xi_b} d\phi_2]^2 + \frac{\rho^2 dr^2}{\Delta_r} + \frac{\rho^2 d\theta^2}{\Delta_\theta} \cr
       &&+\frac{1+g^2 r^2}{r^2 \rho^2} [a\, b\, dt - \frac{b (r^2 +a^2)\sin^2 \theta}{\Xi_a} d\phi_1 - \frac{a(r^2 +b^2) \cos^2\theta}{\Xi_b} d\phi_2]^2\,,
\eea
where
\bea
\Delta_r &=& \fft1{r^2} (r^2 +a^2)(r^2 +b^2)(1+g^2 r^2) -2m\,, \qquad
\Delta_\theta = 1- a^2 g^2 \cos^2\theta- b^2 g^2 \sin^2\theta\,, \cr
\rho^2 &=& r^2 +a^2 \cos^2\theta + b^2 \sin^2\theta \,,\qquad
\Xi_a = 1 - a^2 g^2 \,, \qquad \Xi_b \equiv 1 - b^2 g^2 \,.
\eea
The metric satisfies $R_{\mu \nu} = - 4 g^2 g_{\mu\nu}$.
The mass and angular momenta are \cite{Gibbons:2004ai}:
\bea
M=\frac{\pi m (2\Xi_a + 2\Xi_b - \Xi_a \Xi_b)}{4\Xi_a^2 \Xi_b^2}\,, \qquad J_a=\frac{\pi \,m\, a}{2\Xi_a^2 \Xi_b} \,, \qquad J_b = \frac{\pi \, m \, b}{2\Xi_a \Xi_b^2}\,,
\eea
And Riemann tensor squared is
\be
\text{Riem}^2 = 40 g^4 + \frac{96 m^2 (3\rho^2 - 4 r^2)(\rho^2 - 4 r^2 )}{\rho^{12}}\,.
\ee

We can take the BPS limit by setting
\bea
&&a=\fft1g(1- \fft12 \alpha^2 g^2 \epsilon) \,, \qquad b= \fft1g ( 1- \fft12 \beta^2 g^2 \epsilon)\,, \cr
&&r^2 = -\fft1{g^2}(1- \tilde{r}^2 g^2 \epsilon) \,, \qquad m = g^2 \tilde{m} \epsilon^3\,,
\eea
and sending $\epsilon \rightarrow 0$. The metric becomes
\bea
ds_5^2=&& -[dt + \frac{(\alpha^2 -\tilde{r}^2)\sin^2\theta}{\alpha^2 g} d\phi_1 + \frac{(\beta^2-\tilde{r}^2)\cos^2\theta}{\beta^2 g} d\phi_2]^2 + \frac{\tilde{\rho}^2}{\tilde{\Delta}_\theta} d\theta^2 + \frac{\tilde{\rho}^2}{\tilde{\Delta}_r} d\tilde{r}^2 \cr
   &&+ \frac{\tilde{\Delta}_r \tilde{r}^2 }{\tilde{\rho}^2}(\frac{\sin^2\theta}{\alpha^2 g^2} d\phi_1 + \frac{\cos^2 \theta}{\beta^2 g^2} d\phi_2)^2 + \frac{\tilde{\Delta}_\theta \sin^2\theta \cos^2\theta}{\tilde{\rho}^2}(\frac{\alpha^2 -\tilde{r}^2}{\alpha^2 g^2} d\phi_1 - \frac{\beta^2- \tilde{r}^2}{\beta^2 g^2} d\phi_2)^2 \,,
\eea
where
\bea
\tilde{\Delta}_r &=& \frac{g^2 \tilde{r}^2 (\alpha^2 - \tilde{r}^2)(\beta^2 -\tilde{r}^2)+ 2 \tilde{m}}{\tilde{r}^2}\,, \cr
\tilde{\Delta}_\theta &=& g^2(\alpha^2 \cos^2\theta + \beta^2 \sin^2\theta) \,,\cr
\tilde{\rho}^2 &=& \tilde{r}^2 - \alpha^2 \cos^2\theta- \beta^2 \sin^2\theta \,,\label{d5metrot}
\eea
(An analogous scaling procedure was
used for five-dimensional Kerr-AdS with equal angular momenta in
\cite{Cvetic:2005nc}.)
The metric is a constant time bundle over a four-dimensional Einstein-K\"ahler
space.  The mass and angular momenta become
\bea
\tilde{M} = \frac{\pi  \, \tilde{m} (\alpha^2 + \beta^2)}{2 g^4 \alpha^4 \beta^4}\,, \qquad \tilde{J}_a = \frac{\pi\, \tilde{m}}{2 g^5 \alpha^4 \beta^2} \,,\qquad \tilde{J}_b = \frac{\pi\, \tilde{m}}{2 g^5 \alpha^2 \beta^4} \,,
\eea
satisfying the BPS condition
\be
\tilde M=g\tilde J_a + g \tilde J_b\,.
\ee
The Riemann tensor squared is
\be
\text{Riem}^2 = 40 g^4 + \frac{1536 \tilde{m}^2}{\tilde{\rho}^{12}} \,.
\ee
The metric has a power-law curvature singularity at $\tilde \rho=0$.  For positive $\tilde m$, the singularity is naked. However, when $\tilde m$ is negative, there exist a Euclidean Killing horizon at $r=r_0$ where $\tilde \dlt_r (r_0)=0$.  The absence of the conic singularity associated with the degenerate cycles at $\tilde r=r_0$, $\theta=0$ and $\theta=\pi/2$ requires that the Killing vectors
\bea
\theta=0:&&\ell_1 = \frac{\del}{\del \phi_1} \,, \cr
\theta=\ft{\pi}2:&&\ell_2 = \frac{\del}{\del \phi_2} \,,\cr
\tilde r=r_0:&& \ell = \fft{1}{\kappa}\Big(\frac{\del}{\del t} + \frac{g \alpha^2}{r_0^2 -\alpha^2}\frac{\del}{\del\phi} + \frac{g \beta^2}{r_0^2- \beta^2} \frac{\del}{\del \phi_2}\Big) \,,
\eea
must all generate $2\pi$ period.  Here the Euclidean surface gravity $\kappa$ on the Killing horizon is
\be
\kappa =  \frac{g(3 r_0^4 - 2 (\alpha^2 + \beta^2)r_0^2 + \alpha^2 \beta^2)}{(\alpha^2 -r_0^2)(\beta^2 - r_0^2)}\,.
\ee
It is worth pointing out that the metric (\ref{d5metrot}) is written in the asymptotically rotating frame.  We can make a coordinate transformation $\phi_i\rightarrow \phi_i + g t$ such that the metric becomes non-rotating asymptotically. This implies that
\be
\ell\rightarrow \ell = \fft{1}{\kappa}\Big(\frac{\del}{\del t} + \frac{g r_0^2}{r_0^2 -\alpha^2}\frac{\del}{\del\phi} + \frac{g r_0^2}{r_0^2- \beta^2} \frac{\del}{\del \phi_2}\Big)\,.
\ee
Defining $\ell_0=g^{-1} \partial_t$, we see that the Killing vectors must satisfy the linear relation
\be
p \ell_0 = q \ell + q_1 \ell_1 + q_2\ell_2\,,
\ee
with
\be
p=q + q_1 + q_2\,.
\ee
Consistency requires that $(p,q,q_1,q_2)$ are coprime integers, and consequently $\Delta t=2\pi$. The integration constants can expressed in terms of two rational numbers $(p/q_1,p/q_2)$:
\be
\alpha^2 = \big(1 + \fft{p}{q_1}\big) r_0^2\,,\qquad \beta^2 = \big(1 + \fft{p}{q_2}\big) r_0^2\,.
\ee
The mass and angular momenta are completely discretised, given by
\bea
&&M=-\frac{\pi  p^2 \left(p q_1+p q_2+2 q_2 q_1\right)}{4 g^2 \left(p+q_1\right){}^2 \left(p+q_2\right){}^2}\,,\cr
&&
J_a=-\frac{\pi  p^2 q_1}{4 g^3 \left(p+q_1\right){}^2 \left(p+q_2\right)}\,,\qquad
J_b=-\frac{\pi  p^2 q_2}{4 g^3 \left(p+q_1\right) \left(p+q_2\right){}^2}\,.
\eea

\subsection{$D=2n+1$}

The Kerr-AdS metric in $D=2n+1$ dimensions is given by \cite{Gibbons:2004uw,Gibbons:2004js}
\bea
ds_D^2 &=& - W (1+g^2 r^2)dt^2 + \frac{2 m}{U}(W dt - \sum_{i=1}^n \frac{a_i \mu_i^2}{\Xi_i} d\vp_i)^2 +\sum_{i=1}^n \frac{r^2 + a_i^2}{\Xi_i} \mu_i^2 d\vp_i^2 \cr
&&+ \frac{U}{V-2m} dr^2 + \sum_{i=1}^n \frac{r^2 + a_i^2}{\Xi_i} d\mu_i^2
 -\frac{g^2}{W (1+ g^2 r^2)}(\sum_{i=1}^n\frac{r^2 + a_i^2}{\Xi_i} \mu_i d\mu_i)^2 \,,
\eea
where
\bea
&&W \equiv \sum_{i=1}^n \frac{\mu_i^2}{\Xi_i} \,, \qquad U\equiv \sum_{i=1}^n \frac{mu_i^2}{r^2 + a_i^2} \prod_{j=1}^n(r^2 + a_j^2) \,,\cr
&&V \equiv r^{-2} (1+ g^2 r^2) \prod_{j=1}^n (r^2+a_j^2) \,, \qquad \Xi_i \equiv 1 - a_i^2 g^2 \,,\qquad \sum_{i=1}^n \mu_i^2 = 1 \,.
\eea
They satisfy $R_{\mu\nu} = - (D-1) g^2 g_{\mu\nu}$. The mass and angular momenta are \cite{Gibbons:2004ai}
\be
M = \fft{m {\cal A}_{D-2}}{4 \pi (\prod_{j} \Xi_j)}(\sum_{i=1}^n\frac{1}{\Xi_i} - \fft12)\,,\qquad J_i = \fft{m a_i {\cal A}_{D-2}}{4 \pi \Xi_i (\prod_j \Xi_j)}\,,
\ee
The metric is non-rotating at asymptotic infinity.  We take the following transformation,
\be
\si_i = \vp_i - a_i g^2 t\,,
\ee
so that $g_{tt}\rightarrow -1$ at asymptotic infinity. We now take the BPS limit by setting
\bea
a_i = \fft1g(1- \fft12 b_i^2 g^2 \ep) \,, \qquad r^2 = -\fft1{g^2}(1-y^2g^2 \ep)\,,\qquad m=g^2 \tilde{m} \ep^{n+1}\,,
\eea
and sending $\epsilon \rightarrow 0$. The metric becomes
\bea
ds^2_d = &&-\Big(dt - \fft1g\sum_{i=1}^n\frac{y^2-b_i^2}{b_i^2} \mu_i^2 d\si_i\Big)^2 + \frac{\dlt_\si y^2 dy^2}{\dlt_y} + \frac{\tilde{m}}{g^4 \dlt_\si}(\sum_{i=1}^n \frac{\mu_i^2}{b_i^2} d\si_i)^2 \cr
         &&+\fft1{g^2} [(\sum_{i=1}^n \frac{y^2-b_i^2}{b_i^2} \mu_i^2 d\si_i)^2+\sum_{i=1}^n\frac{y^2-b_i^2}{b_i^2} \mu_i^2 d\si_i^2] \cr
         &&+ \sum_{i=1}^n \frac{y^2 -b_i^2}{b_i^2 g^2}d\mu_i^2 - \frac{1}{ g^2 \dlt_\mu} (\sum_{i=1}^n\frac{y^2 -b_i^2}{b_i^2} \mu_i d\mu_i)^2\,,
\eea
where
\bea
&&\dlt_\mu = y^2 \sum_{i=1}^n \frac{\mu_i^2}{b_i^2}  \,, \qquad \dlt_\si = (\sum_{i=1}^n \frac{\mu_i^2}{y^2-b_i^2}) \prod_{j=1}^n(y^2 -b_j^2)\,, \cr &&\dlt_y = \tilde{m} + g^2 y^2 \prod_{i=1}^n (y^2 -b_i^2)  \,.
\eea
The metric is again constant time bundle over $D=2n$ space, indicating that the solution admits a Killing spinor. The mass and angular momenta become
\be
\tilde M= \frac{\tilde{m} {\cal A}_{D-2}}{4 \pi g^{2n} (\prod_j b_j^2)} \sum_{i=1}^n \frac{1}{b_i^2} \,, \qquad \tilde{J}_i = \frac{\tilde{m} {\cal A}_{D-2}}{4 \pi g^{2n+1} (\prod_j b_j^2)} \cdot \frac{1}{b_i^2} \,.
\ee
satisfying the BPS condition
\be
\tilde M = g \sum_{i=1}^n \tilde J_i \,.
\ee
The metric has a power-law curvature singularity at $\dlt_\psi=0$.  The singularity is naked for positive $\tilde m$, but outside the Euclidean Killing horizon $y_0$ with $\Delta_y=0$.  The Killing vectors associated with the degenerated null surfaces are
\bea
\ell &=& \frac{1}{\kappa}(\frac{\del}{\del t}+ \sum_{i=1}^n \frac{g b_i^2}{y_0^2 -b_i^2} \frac{\del}{\del \si_i})\,, \qquad (y=y_0) \cr
\ell_k &=& \frac{\del}{\del \si_k} \,,\qquad (\mu_k = 0\,, \; k = 1 \cdots n) \,.
\eea
Here the surface gravity $\kappa$ on the horizon is
\be
\kappa = g\Big( 1+ \sum_{i=1}^n \frac{y_0^2}{y_0^2 - b_i^2}  \Big)  \,.
\ee
Making a coordinate transformation $\phi_i\rightarrow \phi_i + gt$, we find that the Killing vector $\ell$ becomes
\be
\ell\rightarrow \ell = \frac{1}{\kappa}(\frac{\del}{\del t}+ \sum_{i=1}^n \frac{g y_0^2}{y_0^2 -b_i^2} \frac{\del}{\del \si_i})\,.
\ee
It follows that the Killing vectors satisfy
\be
p \ell_0= q \ell + \sum_{i=1}^n q_i \ell_i\,,\qquad
\hbox{with}\qquad p=q + \sum_{i=1}^n q_i\,.
\ee
As in the previous $D=5$ case, consistency requires that $\Delta t=2\pi$.

We can now expressed the $n$ integration constant $b_i$ as
\be
b_i^2 = \big(1 + \fft{p}{q_i}\big) r_0^2\,.
\ee
The mass and charges are completely discretised, given by
\bea
M &=& - \fft{{\cal A}_{2n-1}}{4\pi g^{2n-2}} \Big(\prod_i \fft{p}{p + q_i}\Big) \sum_i \fft{q_i}{p+q_i}\,,\cr
J_i &=& - \fft{{\cal A}_{2n-1}}{4\pi g^{2n-1}} \Big(\prod_j \fft{p}{p + q_j}\Big) \fft{q_i}{p+q_i}\,.
\eea

\subsection{$D=2n+2$}

The Kerr-AdS metric in $D=2n+2$ dimensions is given by \cite{Gibbons:2004uw,Gibbons:2004js}
\bea
ds_D^2 &=& - W (1+g^2 r^2)dt^2 + \frac{2 m}{U}(W dt - \sum_{i=1}^n \frac{a_i \mu_i^2}{\Xi_i} d\vp_i)^2 +\sum_{i=1}^n \frac{r^2 + a_i^2}{\Xi_i} \mu_i^2 d\vp_i^2\cr
 &&+ \frac{U}{V-2m} dr^2 + \sum_{i=0}^n \frac{r^2 + a_i^2}{\Xi_i} d\mu_i^2
  -\frac{g^2}{W (1+ g^2 r^2)}(\sum_{i=0}^n\frac{r^2 + a_i^2}{\Xi_i} \mu_i d\mu_i)^2 \,,
\eea
where $a_0=0$ and
\bea
&&W \equiv \sum_{i=0}^n \frac{\mu_i^2}{\Xi_i} \,, \qquad U\equiv \sum_{i=0}^n \frac{\mu_i^2}{r^2 + a_i^2} \prod_{j=1}^n(r^2 + a_j^2) \,,\cr
&&V \equiv r^{-2} (1+ g^2 r^2) \prod_{j=1}^n (r^2+a_j^2) \,, \qquad \Xi_i \equiv 1 - a_i^2 g^2 \,,\qquad \sum_{i=0}^n \mu_i^2 = 1 \,.
\eea
They satisfy $R_{\mu\nu} = - (D-1) g^2 g_{\mu\nu}$. The mass and angular momenta are
\be
M = \fft{m {\cal A}_{D-2}}{4 \pi (\prod_{j} \Xi_j)}\sum_{i=1}^n\frac{1}{\Xi_i}\,,\qquad J_i = \fft{m a_i {\cal A}_{D-2}}{4 \pi \Xi_i (\prod_j \Xi_j)}\,.
\ee
As in the odd-dimensional case, we first make the coordinate transformation
\be
\si_i = \vp_i - a_i g^2 t\,.
\ee
The BPS condition $M=g\sum_i J_i$ can be satisfied by setting
\bea
a_i = \fft1g (1- \ft12 b_i^2 g^2 \ep) \,,
\qquad r^2 = -\fft1{g^2}(1-y^2g^2 \ep)\,,
\qquad m=g^2 \tilde{m} \ep^{n+1}\,,
\eea
and sending $\epsilon \rightarrow 0$. We then make the further transformations
\bea
\theta= i \tilde{\theta} \,, \qquad \mu_0 = \sin\theta \,,\qquad
\mu_i = \cos\theta \, \tilde\mu_i \,,(i=1,\cdots,n)\,,
\eea
with $\sum\tilde \mu_i^2=1$. The $(2n+2)$-dimensional metric can now be
expressed as a foliation of a $(2n+1)$-dimensional BPS time machine
\bea
ds^2_{2n+2} = g^{-2} d\tilde{\theta}^2 + \cosh^2\tilde{\theta}
\; ds^2_{2n+1} \,.
\eea

So far, we have considered the general class of BPS Kerr-AdS time machines
in both odd and even dimensions, with generic but non-vanishing angular
momenta.  When some subset of the angular momenta vanish, the BPS limits
also exist.  For a general Kerr-AdS black hole in $D$ dimensions, if there
are $p$ non-vanishing angular momenta, the resulting BPS time machine
metric takes the form
\be
ds^2_D = g^{-2} d\tilde \theta^2 + \cosh^2\tilde\theta\, ds_{2p+1}^2 +
\sinh^2\tilde\theta\, d\Omega_{D-2p-2}^2\,,
\ee
where $ds_{2p+1}^2$ is the metric for the BPS time machine in $(2p+1)$ dimensions.

\section{Further comments in $D=3$}

   The solutions we gave in section \ref{equaloddsec} specialise
to $D=3$ dimensions if we set $n=1$. It is instructive to compare this with the BTZ black
hole solution \cite{Banados:1992wn} since they are, of course, necessarily locally
equivalent, both being locally just AdS$_3$.

 The BTZ black hole is given by the metric \cite{Banados:1992wn}
\bea
ds^2 &=& -N^2 dt^2 + \frac{d\rho^2}{N^2}+
            \rho^2(d\phi - \frac{J}{2 \rho^2} dt)^2 \,, \cr
N^2 &=& -M + g^2 \rho^2 + \frac{J^2}{4 \rho^2}\,,\label{BTZ}
\eea
and the mass and angular momentum are
\be
M_{\rm BTZ} = g^2(\rho_+^2 + \rho_-^2 ) \,, \qquad J_{\rm BTZ}
   = 2 g \rho_+ \rho_- \,,
\ee
where $\rho_+$ and $\rho_-$ are the radii of the outer and inner horizons.
The BPS limit $M_{\rm BTZ} = g J_{\rm BTZ}$ implies that
$\rho_+ = \rho_- = \rho_0$, and then
\bea
ds^2 &=& -N^2 dt^2 + \frac{d\rho^2}{N^2}+
  \rho^2(d\phi - \frac{g \rho_0^2}{\rho^2} dt)^2 \,, \cr
N^2 &=& \frac{g^2 (\rho^2 - \rho_0^2)^2}{\rho^2}\,. \label{BTZsusy}
\eea

  The rotating $D=3$ black hole following from (\ref{oddequalrot2}) by
setting $n=1$ is
\bea
ds_{3}^2 &=& \fft{dr^2}{f} - \fft{f}{W} dt^2 + r^2 W
(d\phi - \fft{\sqrt{\mu\nu}}{r^{2}+\nu}\, dt)^2 \,,\cr
f &=& (1+g^2 r^2) W - \mu\,,\qquad W=1 + \fft{\nu}{r^{2}}\,.\label{our1}
\eea
Making the coordinate redefinition
\be
r^2=\rho^2 -\nu\,,\label{rtorho}
\ee
we see that (\ref{our1}) becomes
\bea
ds^2 &=& - h dt^2 + \frac{d\rho^2}{h} + \rho^2
(d\phi - \frac{\sqrt{\mu\nu}}{\rho^2} dt)^2 \,,\cr
h &=& g^2 \rho^2 + 1 -(g^2 \nu + \mu) + \frac{\mu \nu}{\rho^2} \,.\label{our2}
\eea
According to our general formulae (\ref{odddmj}),
the mass and angular momentum are given by
\be
M = \mu + g^2\nu \,, \qquad J = 2 \sqrt{\mu\nu} \,.
\ee
Comparing (\ref{our2}) with the BTZ black hole metric (\ref{BTZ}), we
see that they match completely, with
\be
M_{\rm BTZ}= M-1\,,\qquad J_{\rm BTZ}= J\,.\label{btznewsol}
\ee

   The above relations between the mass and angular momentum however
give very different physical interpretations of the seemingly equivalent
solution.  In particular, they lead to very different
BPS conditions
\be
M=g J\,,\qquad \hbox{or}\qquad M_{\rm BTZ} = g J_{\rm BTZ}\,.
\ee
At the first sight, it would seem surprising if both conditions were to
lead to well-defined Killing spinors.

     Before solving the Killing spinor equations, we note that the
vacuum for the BTZ metric with $M_{\rm BTZ}=0=J_{\rm BTZ}$ is AdS$_3$
in planar coordinates, whilst the vacuum for our metric, defined by
$M=0=J$, yields AdS$_3$ in global coordinates:
\bea
M_{\rm BTZ}=0= J_{\rm BTZ}:&&
ds^2= -g^2 \rho^2 dt^2 + \fft{d\rho^2}{g^2 \rho^2} + \rho^2 d\phi^2\,,\cr
M=0=J:&&
ds^2=-(g^2 \rho^2+1) dt^2 + \fft{d\rho^2}{g^2 \rho^2+1} + \rho^2 d\phi^2\,.
\eea

   To derive the Killing spinors, it is convenient to choose the
vielbein basis
\be
e^0=-N dt\,,\qquad e^1 =
\fft{d\rho}{N}\,,\qquad e^2=\rho (d\phi -\Omega dt)\,,\quad
\hbox{with}\quad \Omega = \fft{J}{2\rho^2}\,.
\ee
Note that we use $(0,1,2)$ to denote tangent indices and
$(t,\rho,\psi)$ to denote spacetime indices.
The spinor-covariant exterior derivative
$D = d + \fft14 \omega^{ab}\gamma_{ab}$ is
\bea
D &=& d \otimes \oneone + e^0 \Big(\fft{N'}{2} \gamma_{01} +
\fft{\rho \Omega'}{4} \gamma_{12} \Big) +
e^1 \fft{\rho \Omega'}{4} \gamma_{02} - e^2 \Big(\fft{N}{2 \rho}
\gamma_{12} - \fft{\rho \Omega'}{4} \gamma_{01} \Big)\,, \cr
d &=& e^0 \Big(\fft1N \fft{\del}{\del t} +
 \fft{\Omega}{N} \fft{\del}{\del \phi} \Big) +
e^1 N \fft{\del}{\del \rho} + e^2 \fft1{\rho} \fft{\del}{\del \phi} \,,
\eea
where the Dirac matrices are defined in (\ref{2Dirac}).
We find that the two-component Killing spinor is given by
\be
 \zeta = e^{\ft12 \Delta (g t +\phi )}\,
   \begin{pmatrix} \zeta_+(\rho)\\
                   \zeta_{-}(\rho) \end{pmatrix}\,.
\ee
where $(\zeta_+,\zeta_-)$ satisfy the constraints
\bea
0&=&2 \rho \Big(J\pm 2 \rho \Delta - g \rho \Big)\zeta'_\pm +
\big(J+ 2 g \rho^2\big) \zeta_\pm  \,, \cr
\frac{\zeta_+}{\zeta_-} &=& -
\frac{J^2 - 4 M \rho^2 + 4 g^2 \rho^4}{2\rho(2\rho\Delta+ 2g \rho^2-J)}\,,
\eea
and the exponent $\Delta$ is given by
\be
\Delta = \sqrt{M_{\rm BTZ} - g J_{\rm BTZ}}\,,
\qquad\hbox{or equivalently,}\qquad
\Delta = \sqrt{M - g J-1}\,.
\ee

   The situation becomes clear now with the explicit Killing spinor solutions.
Owing to the fact that the three-dimensional metric is locally AdS$_3$,
the Killing spinors exist locally for all mass and charge, regardless
whether they satisfy the BPS conditions or not.  For the BTZ black holes
$M_{\rm BTZ} > g J_{\rm BTZ}$, the local Killing spinor has real
exponential dependence on the $\phi$ coordinate.  However, since $\phi$
must be periodic in order for the solution to describe a black hole,
as opposed to AdS$_3$, the Killing spinor can only be well defined when
$M_{\rm BTZ}=g J_{\rm BTZ}$, implying that $\Delta$ becomes zero and so
the Killing spinor no longer depends on $\phi$.
Note that for the Killing vector $K=\partial_t + g\partial_\phi$, we have
\be
g(K,K)=\Delta^2\ge 0\,.
\ee
Thus, the Killing vector associated with the Killing spinor is null for the
supersymmetric BTZ black hole, corresponding to $\Delta =0$.

   This is not the only way to achieve the supersymmetry, however.
We can instead impose $M=gJ$, corresponding to $M_{\rm BTZ}-gJ_{\rm BTZ}=-1$,
in which case, we have
\be
\Delta=\sqrt{-1} = {\rm i}\,,\qquad g(K,K)=\Delta^2=-1\,.
\ee
In this case, the Killing vector is time-like, and the Killing spinor now has
periodic dependence on $\phi$, with the same period as that in the
global AdS$_3$.  The resulting metric with negative mass then leads
to the BPS time machine.

Killing spinors of BTZ black holes were also studied in \cite{Coussaert:1993jp,Colgain:2016lxi}.

\section{Conclusions}

In this paper, we studied the global structure of the Kerr-AdS metrics
in general dimensions, when the mass and angular momenta satisfy the
BPS condition (\ref{BPScons}).  In odd dimensions with equal angular
momenta, we construct explicitly the Killing spinors.

For positive mass, the solutions have naked power-law curvature singularities
with no horizon to cloak them. For negative mass, the BPS solutions describe
smooth spacetime configurations that are called time machines.
These smooth spacetime configurations are purely gravitational
and there is no matter energy-momentum tensor source at all.
The completeness of the spacetime requires that the asymptotic
Lorentzian time coordinate be periodic, with precisely the same time
period as that of the AdS hyperboloid in which the solutions are immersed.
Furthermore, the mass and angular momenta become discretised.  The Killing
spinors are periodic in time, with a period that is consistent with the
global structure of the time machines.  Thus in cases where they
solutions can be embedded in gauged
supergravities, they are supersymmetric.

In the AdS/CFT correspondence, the time coordinate in both the global or
the planar AdS spacetime is taken to lie on the real line, describing
an infinite covering of the AdS hyperboloid in the global case. In this
case, the BPS time machines constructed in this paper would all have a conical
singularity at the Euclidean Killing horizon.  However, if we consider
the asymptotic AdS as being the strict AdS hyperboloid in $R^{2,D-2}$,
the time machines described in this paper are precisely consistent with
the boundary conditions. The breaking of the time translational symmetry in
our BPS and the general non-BPS \cite{Feng:2016dbw} Kerr-AdS time machines is
reminiscent of the time crystals proposed by Wilczek \cite{Wilczek:2012jt}.

\section*{Acknowledgement}

M.C.~and C.N.P.~are grateful to the Physics Department of Beijing Normal University for hospitality in the beginning stage of this work. M.C.~and H.L.~are grateful to the Mitchell Institute for hospitality
during the completion of this work.  We thank Gary Gibbons and Ergin Sezgin
for useful conversations.
M.C.~is supported in part by DOE Grant Award de-sc0013528, the Fay R. and Eugene L. Langberg Endowed Chair and the Slovenian Research Agency (ARRS) (M.C.).
W.-J.G.~and H.L.~are supported in part by NSFC grants No.~11475024,
No.~11175269 and No.~11235003.
C.N.P.~is supported in part by DOE grant DE-FG02-13ER42020.

\appendix

\section{$\CP^n$ and gauge-covariantly constant spinor}

   Here we make use of the iterative construction of $\CP^n$ in terms of
$\CP^{n-1}$ that was obtained in \cite{Hoxha:2000jf}, in order to give an
explicit
iterative construction of the gauge-covariantly constant spinor that we
employed in the construction of the Killng spinor in the
previous section.  As was
shown in \cite{Hoxha:2000jf}, the Fubini-Study metric $d\Sigma_n^2$ on
$\CP^n$ can be
written in terms of the Fubini-Study metric $d\Sigma_{n-1}^2$
on $\CP^{n-1}$ as
follows:
\be
d\Sigma_n^2 = d\chi^2 + \sin^2\chi\, \cos^2\chi\, (d\psi+ \wtd A)^2 +
  \sin^2\chi\, d\Sigma_{n-1}^2\,,\label{nfromn-1}
\ee
where $\wtd J=\ft12 d\wtd A$ is the K\"ahler form of $\CP^{n-1}$.  The K\"ahler
form of $\CP^n$ is given by $J=\ft12 dA$, where
\be
A= \sin^2\chi\, (d\psi + \wtd A)\,.
\ee

   We define the vielbein $e^a$ for $\CP^n$, with\footnote{Note that although
we are using a 0 index here, in this section it refers to a Euclidean
direction not a time direction.  This section is intended
to be self-contained, and not all notation or symbols used here are the same
as in the previous section.}
\be
e^0=d\chi\,,\qquad e^1= \sin\chi\, \cos\chi\, (d\psi+\wtd A)\,,\qquad
  e^i= \sin\chi\, \td e^i\,,
\ee
where $\td e^i$ is a vielbein for $\CP^{n-1}$.  The inverse vielbein is then
given by
\be
E_0= \fft{\del}{\del\chi}\,,\qquad
E_1= \fft1{\sin\chi\,\cos\chi}\, \fft{\del}{\del\psi}\,,\qquad
E_i= \fft1{\sin\chi}\, \Big(\wtd E_i - \wtd A_i\, \fft{\del}{\del\psi}\Big)\,.
\ee
A straightforward calculation shows that the spinor-covariant exterior
derivative $D=d+\ft14 \omega^{ab}\, \Gamma_{ab}$ on $\CP^n$ is given by
\be
D=d + +\ft14\td\omega^{ij}\, \Gamma_{ij} -e^1\,
 \Big(\cot2\chi\,\Gamma_{01} +
      \ft14 \cot\chi\, \wtd J^{ij}\, \Gamma_{ij}\Big) -
    \ft12 e^i \,\cot\chi\, \Big(\Gamma_{0i} + \wtd J_i{}^j\, \Gamma_{1j}
\Big)\,.\label{ncov}
\ee

   Decomposing the $2n$-dimensional Dirac matrices $\Gamma_a$ for $\CP^n$ as
\be
\Gamma_0= \sigma_2\otimes \wtd\Gamma_*\,,\qquad
  \Gamma_1= \sigma_1\otimes \wtd\Gamma_*\,,\qquad
    \Gamma_i= \oneone\otimes \wtd\Gamma_i\,,
\ee
where $\wtd\Gamma_i$ are the $(2n-2)$-dimensional Dirac matriaces for
$\CP^{n-1}$, it can be seen that the spinor-covariant exterior derivative
(\ref{ncov}) can be written as
\bea
D &=& \oneone\otimes \wtd D + e^0\, \fft{\del}{\del\chi} +
    e^1\, \fft1{\sin\chi\,\cos\chi}\, \fft{\del}{\del\psi} -
   e^i\, \fft1{\sin\chi}\, \wtd A_i\, \fft{\del}{\del\psi}
  + \im\, e^1 \cot2\chi \, \sigma_3\otimes\oneone\nn\\
&& -
  \ft14 e^1\, \cot\chi\, \wtd J^{ij}\, \oneone\otimes \Gamma_{ij}-
  \ft12 e^i\, \cot\chi\, \Big(\sigma_2\otimes \wtd\Gamma_*\wtd\Gamma_i
  + \wtd J_i{}^j\, \sigma_1\otimes \wtd\Gamma_*\wtd\Gamma_j\Big)\,,
\eea
where $\wtd D=\td d+ \ft14 \td\omega^{ij}\, \wtd\Gamma_{ij}$ is the
spinor-covariant exterior derivative on $\CP^{n-1}$.

   Assuming that the $\CP^{n-1}$ admits a gauge-covariantly constant
spinor $\td\xi$ satisfying
\be
\wtd D + \ft{\im}{2}\, n\, \wtd A\, \td\xi=0\,,\qquad
    \wtd J_i{}j\, \wtd\Gamma_{j}\, \td\xi =-\im\, \wtd\Gamma_j\, \td\xi\,,
\qquad
\wtd\Gamma_*\, \td\xi= \td\xi
\ee
(the middle equation also implies $\wtd J^{ij}\, \wtd \Gamma_{ij}\, \td\xi=
 -2\im\, (n-1)\, \td\xi$),
it then follows that $\CP^n$ admits a gauge-covariantly constant spinor
$\xi=\nu\otimes \td\xi$ satisfying
\be
D\xi + \ft{\im}{2}\, (n+1)\, A\, \xi=0\,,\qquad
J_a{}_b\, \Gamma_{ab}\, \xi= -\im\, \Gamma_a\, \xi\,,\qquad
\Gamma_*\, \xi=\xi\,,
\ee
where $\Gamma_*=\sigma_3\otimes \wtd\Gamma_*$ is the chirality operator
on $\CP^n$, and
where the 2-component spinor $\nu$ has $\psi$ dependence $e^{-\ft{\im}{2}\,
n\, \psi}$, it depends on no other coordinates, and it obeys $\sigma_3\,
\nu=\nu$.  In other words, the gauge-covariantly constant spinor on
$\CP^n$ can be taken to be
\be
\xi = e^{-\ft{\im}{2}\, n\, \psi}\, \begin{pmatrix}1 \\ 0\end{pmatrix}
\otimes \td\xi\,.
\ee
It also follows that that $\xi$ obeys $J^{ab}\, \Gamma_{ab}\, \xi
= -2\im\, n\, \xi$.

   If we denote the fibre coordinate $\psi$ in the construction
(\ref{nfromn-1}) of $\CP^n$ from $\CP^{n-1}$ by $\psi_n$ we therefore
have an iterative construction of the gauge-covariantly constant spinor:
\be
\xi(\CP^n)= e^{-\ft{\im}{2}\, n\, \psi_n}\, \begin{pmatrix}1 \\ 0\end{pmatrix}
\otimes \xi(\CP^{n-1})\,.
\ee
An almost trivial calculation confirms that for $n=1$ the spinor
\be
\xi(\CP^1)= e^{-\ft{\im}{2}\, \psi_1}\,
\begin{pmatrix}1 \\ 0\end{pmatrix}
\ee
indeed satisfies all the properties assumed above, and so by induction we
arrive at the expression
\be
\xi(\CP^n) = \exp\Big[ -\ft{\im}{2}\, \sum_{p=1}^n p\, \psi_p\Big]\,
 \underbrace{\begin{pmatrix}1 \\ 0\end{pmatrix} \otimes
  \begin{pmatrix}1 \\ 0\end{pmatrix}\otimes \cdots
  \otimes \begin{pmatrix}1 \\ 0\end{pmatrix} }_{n\ \hbox{factors}}
\ee
for the gauge-covariantly constant spinor on $\CP^n$.
(Note that for $n=1$, writing $\chi=\ft12\theta$ and $\psi_1=\phi$ puts
the metric (\ref{nfromn-1}) in the standard form $d\Sigma_1^2=
 \ft14(d\theta^2 + \sin^2\theta\, d\phi^2)$.)

\section{Identities for spinorial square roots}

   In this appendix, we record some basic results for spinors in odd
dimensions, which are related to our discussion about the Killing
vector (\ref{kvplus}) in the time-machine spacetimes.

   In the odd dimension $D=2n+1$, the Fierz identity can be written in the form
\be
\chi\bar\chi= \sum_{p=0}^n \fft{(-1)^{\fft12 p(p-1)}}{2^n\, p!}\,
   \bar\chi \Gamma_{\mu_1\cdots \mu_p}\chi \, \Gamma^{\mu_1\cdots \mu_p}\,,
\label{fierz}
\ee
where $\chi$ is any commuting spinor.  A useful identity is
\be
\Gamma_{\nu_1\cdots\nu_q}\, \Gamma_{\mu_1\cdots\mu_p}\,
  \Gamma^{\nu_1\cdots\nu_q} = c(q,p)\, \Gamma_{\mu_1\cdots\mu_p}\,,
\label{qpGamma}
\ee
where \cite{vanproeyen}
\be
c(q,p)= (-1)^{\fft12 q(q-1)}\, (-1)^{pq}\, q!\,
  \sum_{i=0}^{ {\rm min}(q,p)}\, \binom{p}{i}\, \binom{2n+1-q}{q-i}\, (-1)^i\,,
\label{cqpdef}\,.
\ee

   If we define the tensors
\be
T_{\mu_1\cdots \mu_p} = \bar\chi \Gamma_{\mu_1\cdots \mu_p}\chi\,,
\ee
and their norms
\be
N_{(p)} = T^{\mu_1\cdots\mu_p}\, T_{\mu_1\cdots \mu_p}\,,
\ee
then it is straightforward to see from (\ref{fierz}) and (\ref{qpGamma}) that
these satisfy the set of linear relations
\be
N_{(q)} = \sum_{p=0}^n \fft{(-1)^{\fft12 p(p-1)}}{2^n\, p!}\, c(q,p)\,
   N_{(p)}\,,\qquad 0\le q\le n\,.\label{Nrels}
\ee
(One does not need to consider $q>n$, since $\Gamma_{\mu_1\cdots\mu_p}$ is
proportional to $\Gamma_{\mu_1\cdots \mu_{2n+1-p}}$.)
The equations (\ref{Nrels}) are not, in fact, all linearly independent.  For
example, for the first few cases we find the relations imply:
\bea
D=3:\qquad && N_{(1)} = N_{(0)}\,,\nn\\
D=5:\qquad && N_{(1)}= N_{(0)}\,,\qquad N_{(2)} = -4 N_{(0)}\,,\nn\\
D=7:\qquad && N_{(2)}= - 6 N_{(1)}\,,\qquad N_{(3)}= -42 N_{(0)} +
           24 N_{(1)}\,,\\
D=9: \qquad && N_{(3)}= -36 N_{(0)} + 36 N_{(1)} + 3 N_{(2)}\,,\quad
      N_{(4)} = 216 N_{(0)} + 120 N_{(1)} + 24 N_{(2)}\,,\nn\\
D=11:\qquad && N_{(3)}= -30 N_{(0)} + 30 N_{(1)} + 3 N_{(2)}\,,\quad
   N_{(4)} = 240 N_{(0)} + 120 N_{(1)} + 12 N_{(2)}\,,\nn\\
&& N_{(5)} = 1920 N_{(0)} - 120 N_{(1)} + 60 N_{(2)}\,.\nn
\eea
Only in the first two cases, in $D=3$ and $D=5$ dimensions, we see that
$N_{(1)}$ is simply equal to $N_{(0)}$.  This means that in these two
cases, and only in these cases, one has the relation
\be
  (\bar\chi \Gamma^\mu\chi)(\bar\chi\Gamma_\mu\chi)=
  (\bar\chi\chi)^2\,,\label{N1N0rel}
\ee
where $\chi$ is {\it any} commuting spinor.\footnote{We emphasise that
the spinor $\chi$ here is completely arbitrary, and need not be Majorana.
If one does require $\chi$ to be Majorana, then (\ref{N1N0rel}) will hold
in $D=9$ also, since $C\Gamma_{\mu\nu}$ and $C\Gamma_{\mu\nu\rho}$ are
antisymmetric in $D=9$, so then $N_{(2)}=0$ and $N_{(3)}=0$.}

   The fact that (\ref{N1N0rel}) holds for any commuting spinor in $D=3$
or $D=5$ implies in particular that in these dimensions, any Killing
vector $K^\mu$ that has a spinorial square root, meaning that it can be written
as in terms of a Killing spinor $\eta$ as $K^\mu=\bar\eta\Gamma^\mu \eta$,
will necessarily have constant (negative) norm.

   The Killing vector (\ref{kvplus}) in the BPS time-machine spacetime
has constant and negative norm $K^\mu K_\mu=-1$ in any odd dimension,
and we saw in section \ref{kssec} that it
always has a spinorial square root, as in (\ref{spinorialsqrt}).  In
odd dimensions $D\ge7$, the fact that the norm is constant therefore
depends upon special additional properties of the Killing spinor $\eta$
that would, a priori, not necessarily hold for an arbitrary Killing spinor.

\end{document}